\documentclass[sigconf]{acmart}
\usepackage[utf8]{inputenc}
\usepackage[T1]{fontenc}  
\usepackage[normalem]{ulem}
\usepackage{amsmath}
\usepackage{bbm}
\usepackage{xcolor}

\usepackage{natbib}
\usepackage{graphicx}
\usepackage{subfigure}
\usepackage[skip=-1pt]{caption}
\usepackage{amsmath}
\usepackage{makecell}
\usepackage{amsfonts}
\usepackage{hyperref}
\usepackage{enumitem}
\usepackage{booktabs}
\usepackage{multirow}
\usepackage{mathtools}
\usepackage{multicol}
\usepackage{amsmath, bm}
\usepackage{stmaryrd}
\usepackage{longtable}
\usepackage{bbm}
\DeclareMathOperator*{\argmin}{argmin}
\DeclareMathOperator*{\argmax}{argmax}
\usepackage[ruled,vlined,english,onelanguage]{algorithm2e}

\usepackage{xcolor}
\def\bluelozenge{\mathbin{\color{blue}\blacklozenge}}
\def\redlozenge{\mathbin{\color{purple}\blacktriangle}}

\graphicspath{ {./fig/}  }

\copyrightyear{2022}
\acmYear{2022}
\setcopyright{acmcopyright}\acmConference[CIKM '22]{Proceedings of the 31st ACM International Conference on Information and Knowledge Management}{October 17--21, 2022}{Atlanta, GA, USA}
\acmBooktitle{Proceedings of the 31st ACM International Conference on Information and Knowledge Management (CIKM '22), October 17--21, 2022, Atlanta, GA, USA}
\acmPrice{15.00}
\acmDOI{10.1145/3511808.3557229}
\acmISBN{978-1-4503-9236-5/22/10}
\acmDOI{10.1145/3511808.3557229}

\begin{document}

\title{Adapting Triplet Importance of Implicit Feedback for Personalized Recommendation}

\author{Haolun Wu}
\affiliation{
  \institution{McGill University}
  \city{Montréal}
  \country{Canada}
}
\email{haolun.wu@mail.mcgill.ca}

\author{Chen Ma}
\authornotemark[1]
\affiliation{
  \institution{City University of Hong Kong}
  \country{Hong Kong SAR}
}
\email{chenma@cityu.edu.hk}

\author{Yingxue Zhang}
\affiliation{
  \institution{Huawei Noah's Ark Lab}
  \city{Montréal}
  \country{Canada}
}
\email{yingxue.zhang@huawei.com}

\author{Xue Liu}
\affiliation{
  \institution{McGill University}
  \city{Montréal}
  \country{Canada}
}
\email{xueliu@cs.mcgill.ca}

\author{Ruiming Tang}
\affiliation{
  \institution{Huawei Noah's Ark Lab}
  \city{Shenzhen}
  \country{China}
}
\email{tangruiming@huawei.com}

\author{Mark Coates}
\affiliation{
  \institution{McGill University}
  \city{Montréal}
  \country{Canada}
}
\email{mark.coates@mcgill.ca}

\begin{abstract}

Implicit feedback is frequently used for developing personalized recommendation services due to its ubiquity and accessibility in real-world systems. In order to effectively utilize such information, most research adopts the pairwise ranking method on constructed training triplets $\langle$\textit{user, positive item, negative item}$\rangle$ and aims to distinguish between positive items and negative items for each user. However, most of these methods treat all the training triplets equally, which ignores the subtle difference between different positive or negative items. On the other hand, even though some other works make use of the auxiliary information (e.g., dwell time) of user behaviors to capture this subtle difference, such auxiliary information is hard to obtain. To mitigate the aforementioned problems, we propose a novel training framework named \textbf{T}riplet \textbf{I}mportance \textbf{L}earning (TIL), which adaptively learns the importance score of training triplets. We devise two strategies for the importance score generation and formulate the whole procedure as a bilevel optimization, which does not require any rule-based design. We integrate the proposed training procedure with several Matrix Factorization (MF)- and Graph Neural Network (GNN)-based recommendation models, demonstrating the compatibility of our framework. Via a comparison using three real-world datasets with many state-of-the-art methods, we show that our proposed method outperforms the best existing models by 3-21\% in terms of Recall@$k$ for the top-$k$ recommendation.


\end{abstract}
 \vspace{-5mm}



\maketitle
\renewcommand{\shortauthors}{Haolun Wu et al.}

\section{Introduction}









With the ever-growing volume of online information, Internet users can easily access an increasingly vast number of online products and services. It is also becoming very difficult for users to identify the items that will appeal to them out of a plethora of candidates. To reduce information overload and to satisfy the diverse needs of users, personalized recommendation systems have emerged and they are beginning to play an important role in modern society.

Recommendation systems with implicit feedback are more commonly seen in real-world application scenarios since implicit feedback is easier to collect, compared to explicit feedback~\cite{RendleFGS2009_bpr}. With implicit feedback, only historical interaction data (without users' explicit preferences) are available, such as clicks and downloads. In order to learn the user preference from implicit feedback, one seminal work proposes the Bayesian Personalized Ranking (BPR) loss~\cite{RendleFGS2009_bpr}, which constructs training triplets, $\langle$\textit{user, positive item, negative item}$\rangle$, based on negative sampling and aims to distinguish the positive and negative items for each user. Although the effectiveness of the vanilla BPR training procedure has been demonstrated for a variety of recommendation architectures~\cite{RendleFGS2009_bpr, sun2020_mgcf, he2020_lgcn}, we perceive that it suffers from an important limitation: it treats all constructed triplets equally. This ignores two aspects of real-world implicit feedback: (i) a user often has different preference levels for different items, even though they are all regarded as positives; (ii) failing to interact with a negative item does not necessarily guarantee a user's negative preference regarding this item, especially for new users and new items. Thus, some sampled negative items may be ``false negatives''. Therefore, assigning equal weights to all training triplets leads to sub-optimal embedding learning.



To address the aforementioned problem, several effective approaches have been proposed. They can be grouped into three categories: (1) \textit{negative item sampling}, e.g., AOBPR~\cite{DBLP:conf/wsdm/RendleF14}, WBPR~\cite{DBLP:journals/jmlr/GantnerDFS12}, PRIS~\cite{DBLP:conf/www/Lian0C20}; (2) \textit{positive pair re-weighting}, e.g., TCE and RCE~\cite{DBLP:conf/wsdm/WangF0NC21}, and (3) \textit{using auxiliary information} (e.g., dwell time) to better model the user's preference~\cite{DBLP:conf/recsys/WenYE19, DBLP:conf/sigir/LiuWD10, DBLP:conf/sigir/LuZMWxLLM19, DBLP:conf/recsys/YiHZLR14}. Existing methods in the first two categories fail to learn the importance of training samples at the triplet level. Each method either adjusts the importance of a positive item or adjusts the importance of a negative item. By focusing exclusively on only one component of a triplet, the methods reduce their effectiveness. In order to show this more clearly, we consider the example of a user $u$ who only likes superhero movies. For such a user, $\langle\textit{u}, \textit{Iron Man}, \textit{Spider Man}\rangle$ should have a small weight in order to avoid pushing \textit{Spider Man} too far from the user. By contrast, $\langle\textit{u}, \textit{Iron Man}, \textit{Titanic}\rangle$ should have a larger weight to push \textit{Titanic} further away. Therefore, as illustrated in this example, the importance of the training triplet is not only determined by the item, or the user, or even the user-item pair (e.g., $\langle\textit{u}, \textit{Iron Man}\rangle$), individually. It is the overall relation between the user, the positive item, and the negative item that dictates the significance of a training triplet. This motivates us to directly learn the importance of data at the triplet level during training. Finally, with respect to the third category, the approaches are more restricted because they can only be applied when the appropriate auxiliary information is available, which is often not the case in practical settings.

Directly modeling the importance score at the triplet level leads to several challenges. First, it leads to an enormous learning space if we directly assign a learnable weight to each triplet, with the weight indicating the triplet importance. Successful learning in such a large space imposes a major computational burden. Second, if we adopt a naive, straightforward approach of adding a learnable weight to each triplet in the loss function, then directly optimizing the weights and the embeddings will tend to make the weights converge to zero. This is due to the optimization procedure forcing the weighted loss, which is non-negative, to zero.  

To tackle the aforementioned challenges, we introduce a novel training framework called \textbf{T}riplet \textbf{I}mportance \textbf{L}earning (TIL), which can adjust the contribution of each training triplet. 
In order to address the challenge of an unreasonably large learning space, rather than directly learning the weights assigned to triplets, we design a weight generation function that adapts the triplet importance in a learnable way. 
In our experimental implementation, we model this function using an MLP with one hidden layer. 
It is important to identify suitable inputs for the weight generation function. 
Therefore, we propose two paradigms for designing the input signal of the weight generator. 
The \textit{Uni-Interest Strategy} measures the proximities of the positive item and negative item to the user's principle interest, represented by aggregation of all the items with which the user has interacted. 
The \textit{Multi-Interest Strategy} further represents a user's preference as arising from a mixture of different interests.
We cluster items into different interest categories through an end-to-end clustering method, and then measure the proximities between the user's principle interest and the centers of the interest categories associated by the positive and negative items.


We adopt a bilevel optimization strategy to efficiently learn all the parameters in our framework. In the inner level we strive to optimize the recommendation model parameters, while in the outer level we optimize the MLP parameters for the weight generation function. The whole procedure is trained in an end-to-end fashion and can be applied on any recommendation backbone architecture for learning better embeddings.

To summarize, this work makes the following contributions:
\begin{itemize}[leftmargin=*]

    \item We propose a novel training paradigm, TIL, to improve the training procedure of personalized ranking by adjusting the importance assigned to different triplets. We design a weight generation function to learn triplet importance which avoids enumerating different weights for individual triplets. 
    \item We develop two strategies, Uni-Interest and Multi-Interest, for measuring the proximity of users, positive items, and negative items in constructed training triplets. These strategies provide suitable inputs for the weight generation function. We adopt a self-supervised clustering method to directly learn the cluster center embeddings of items for the Multi-Interest strategy in an end-to-end manner.
    \item  Experiments on three real-world datasets show that TIL significantly outperforms the state-of-the-art methods for the top-$k$ recommendation task. Extensive experiments on four recommendation backbones further demonstrate the compatibility of our method with different architectures.

\end{itemize}

\section{Related Work}

In this section, we discuss related work, focusing on (1) personalized recommendation with implicit feedback from users; and (2) existing approaches towards more effective use of the implicit feedback in recommendation. 

\begin{figure}[t]
    \centering
    \includegraphics[width=\linewidth]{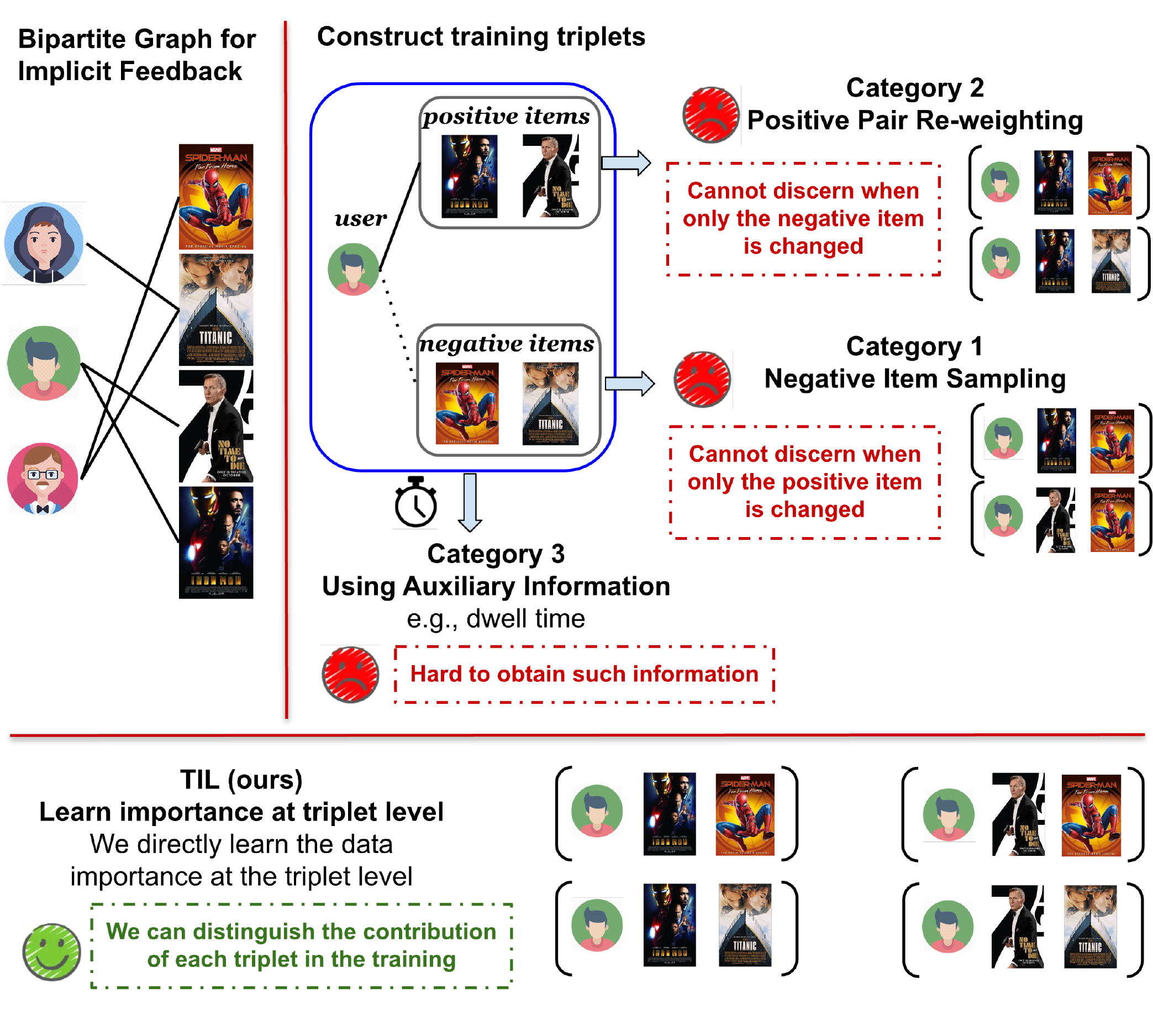}
    \vspace{-5mm}
    \caption{The difference between prior works with our proposed method. This also indicates the novelty of our work.}
    \vspace{-5mm}
\label{pic:summary}
\end{figure}

\subsection{Recommendation with Implicit Feedback}

In many real-world recommendation scenarios, implicit feedback from users \cite{DBLP:conf/cikm/TranLL018}, e.g., browsing or clicking history, is more common than explicit feedback~\cite{DBLP:conf/www/SarwarKKR01,DBLP:conf/icml/SalakhutdinovMH07}. As implicit feedback only provides positive samples, the task of learning from implicit feedback is also called one-class collaborative filtering (OCCF)~\cite{DBLP:conf/icdm/PanZCLLSY08}. Many effective methods have been proposed to tackle this problem. Early approaches either treated all missing data as negative samples \cite{DBLP:conf/icdm/HuKV08}, or sampled negative items from the missing data to learn the pair-wise user preference between positive and negative samples \cite{DBLP:conf/uai/RendleFGS09}. Recent works \cite{DBLP:conf/sigir/HeZKC16,DBLP:conf/www/LiangCMB16} have proposed the assignment of weights to the missing data based on item popularity and user exposure.
Due to their ability to represent non-linear and complex data, (deep) neural networks have been commonly used in the domain of recommendation. 
(Denoising) Autoencoders have been employed to capture the user-item interaction from user implicit feedback by reconstructing the rating matrix~\cite{DBLP:conf/wsdm/WuDZE16,DBLP:conf/wsdm/MaKWWL19}. Recent works using graph neural networks and knowledge graph enhanced recommendation models~\cite{sun2020_mgcf,he2020_lgcn, Ma2021_hyperknow} have also demonstrated their effectiveness in encoding users' preferences from the implicit feedback.
\vspace{-2mm}

\subsection{Effective Use of Implicit Feedback}

How to effectively use the implicit feedback in recommendation is an important research question. We classify the existing works into three categories as demonstrated below and summarize the differences between these works and ours.

\textbf{Category 1: Negative Item Sampling.} Negative item sampling methods focus on making better use of the negative samples so that more informative ones can be highlighted. Rather than trusting all negative items equally, ~\citet{DBLP:journals/jmlr/GantnerDFS12} assumed that popular but uninteracted items are more likely to be real negative items. \citet{DBLP:conf/sigir/ZhangCWY13} drew a set of negative samples from the uniform distribution and selected the item with the largest prediction score. 
AOBPR~\cite{DBLP:conf/wsdm/RendleF14} improved the vanilla BPR with adaptive sampling of negative items, increasing the sampling rate of popular and top ranked items to speed up model convergence. \citet{DBLP:conf/icml/BlancR18} proposed kernel-based sampling, which draws samples proportionally to a quadratic kernel.
Later works designed samplers by constructing hypotheses about user behavior with respect to item properties~\cite{DBLP:journals/jmlr/GantnerDFS12, DBLP:conf/www/DingF0YLJ18}. PRIS~\cite{DBLP:conf/www/Lian0C20} clustered items into several groups based on item representations; each group of unobserved items shared the same probability of being negative based on the prediction score. Other works~\cite{DBLP:conf/www/Lian0C20, DBLP:conf/sigir/YuQ20} proposed negative samplers which consider item popularity and construct clusters of items to better assess user interest.
There also exist several works that used reinforcement learning to discover high-quality negatives~\cite{DBLP:conf/ijcai/DingQ00J19, DBLP:conf/www/WangX000C20}.

\textbf{Category 2: Positive Pair Re-weighting.} The second category of method learns the importance of different data through re-weighting the positive user-item pairs. The re-weighting technique was originally used for data denoising and has been proven effective in various domains~\cite{DBLP:conf/ijcai/Elkan01, DBLP:journals/tnn/KhanHBST18, DBLP:conf/icml/ShenS19, DBLP:conf/nips/ShuXY0ZXM19}. The key idea is to treat those data samples with high loss values as ``noisy'' samples and then assign lower weights to them. Although this strategy has been extensively employed in a variety of domains, there has been limited research conducted on implicit feedback in recommendation. To the best of our knowledge, only TCE/RCE~\cite{DBLP:conf/wsdm/WangF0NC21} has taken the data importance of implicit feedback into consideration. The authors only focused on re-weighting the positive interactions, while omitting the negative ones. Specifically, they devised a re-weighting method to assign lower weights to those positive user-item pairs with large loss values to reduce their influence on the training.


\textbf{Category 3: Using Auxiliary Information.} In addition to the aforementioned categories of approaches, several works have used auxiliary information, such as dwell time, to guide the learning in recommendation with implicit feedback ~\cite{DBLP:conf/wsdm/KimHWZ14, DBLP:conf/sigir/LiuWD10, DBLP:conf/sigir/LuZMWxLLM19, DBLP:conf/recsys/YiHZLR14, LuZM18_between}.

\textbf{Difference and Novelty.} 
The overall difference between prior works and ours is summarized in Fig.~\ref{pic:summary}. The TCE/RCE~\cite{DBLP:conf/wsdm/WangF0NC21} is the existing work most closely related to our proposed method.
The main differences between TCE/RCE and our work are: (1) TCE/RCE~\cite{DBLP:conf/wsdm/WangF0NC21} only considers the noise from potential false-positive items, while we consider the importance of both the positive item and negative item jointly on a triplet level. (2) The weight function in~\cite{DBLP:conf/wsdm/WangF0NC21} for each positive user-item pair is a direct mapping function from the loss value of the pair, which is not learnable and can lead to reinforcement of errors. Our triplet weight generator can easily incorporate a diverse information source such as the proximity of the positive item and negative item to the center of the user’s preference community.

\section{Problem Formulation}
\label{sec:prelimBPR}

The recommendation task considered in this paper takes as input the user implicit feedback. We denote the set of all users and items as $\mathcal{U}$ and $\mathcal{V}$, respectively. 
For each user $ u\in\mathcal{U}$, the user preference data is represented by a set of items he/she has interacted with as $\mathcal{I}_{u}^{+}:=\{i \in \mathcal{I}|\bm{Y}_{u,i}=1\}$ where $\bm{Y} \in \mathbb{R}^{\mathcal{|U|} \times \mathcal{|I|}}$ is the binary implicit feedback rating matrix. We then split $\mathcal{I}_{u}^{+}$ into a training set $\mathcal{S}_{u}^{+}$ and a test set $\mathcal{T}_{u}^{+}$, requiring that  $ \mathcal{S}_{u}^{+} \cup \mathcal{T}_{u}^{+} = \mathcal{I}_{u}^{+} $ and $ \mathcal{S}_{u}^{+} \cap \mathcal{T}_{u}^{+} = \emptyset$. Then the top-$k$ recommendation is formulated as: given the training item set $ \mathcal{S}_{u}^{+} $, and the non-empty test item set $ \mathcal{T}_{u}^{+}$ for user $ u $, the model aims to recommend an ordered set of $k$ items $ \mathcal{X}_{u} $ such that $ |\mathcal{X}_{u}| = k $ and $ \mathcal{X}_{u} \cap \mathcal{S}_{u}^{+} = \emptyset $. Then the recommendation quality is evaluated by some matching metric between $\mathcal{X}_u$ and $\mathcal{T}_u^{+}$ such as Recall@$k$.
\section{Methodology}
In this section, we first introduce Bayesian Personalized Ranking (BPR), a well-established ranking method. Then we present our approach, Triplet Importance Learning (TIL), which assigns varying importance to different triplets during training. We detail two strategies (Uni-Interest and Multi-Interest) for modeling influential factors which can indicate the importance of a training triplet. Thereafter, we describe how to learn the cluster center embeddings of items for the Multi-Interest Strategy in an end-to-end fashion. Finally, we demonstrate our training framework, which uses bilevel optimization to learn the model parameters as well as the triplet importance. 

\subsection{Bayesian Personalized Ranking}
To learn the user preference from implicit feedback, Bayesian Personalized Ranking~\cite{RendleFGS2009_bpr} is a seminal method which has been widely employed. The central idea of BPR is to maximize the ranking of an item that the user has accessed (treated as positive) relative to a randomly sampled item (treated as negative). This goal is achieved via a carefully designed loss function:
\begin{equation}
    \mathcal{L}_{\text{BPR}}(u,i,j;\Theta)=-\mathrm{log} \, \sigma \big(\hat{y}_{ui}(\Theta) - \hat{y}_{uj}(\Theta) \big) \,,
\label{eq:vanilla_BPR}
\end{equation}
where $(u,i,j)$ is a training triplet with a positive item $i$ and a negative item $j$ for user $u$. $\hat{y}_{ui}$ refers to the relevance score between $u$ and $i$, and $\Theta$ is the learnable parameter used to compute $\hat{y}_{ui}$. $\Theta$ includes the user embedding $ \bm{p}_u \in \mathbb{R}^d $ and the item embedding $ \bm{q}_i \in \mathbb{R}^d $, where $d$ is the embedding dimension. 

\subsection{Triplet Importance Enhanced Ranking}
\label{sec:TIL}
Although the vanilla BPR loss has achieved good performance, we argue that treating all training triplets equally may lead to a sub-optimal model for at least two reasons. First, users may have different preference levels for different items, but equal weights cannot distinguish the user preference in a fine-grained manner, making the learned embeddings less tailored to user preference. Second, if the positive item and the randomly sampled negative item are very similar, it is very likely that the user also prefers this sampled item. By optimizing the vanilla BPR loss, the randomly sampled item is pushed away from the user. Therefore, assigning a different level of importance to each training triplet is a promising avenue for improving the performance of a recommendation model.

To incorporate the above intuition, we propose to measure the importance of each triplet using a learnable function. We associate with each triplet a state $\mathbf{s}_{uij}(\Theta)$, and introduce $f(\mathbf{s}_{uij};\Lambda)$, an importance generation function that outputs a weight for each $(u,i,j)$. $\Lambda$ is a set of learnable parameters associated with the function. We then modify the BPR loss as follows:
\begin{equation}
    \mathcal{L}_{\text{I-BPR}}(u,i,j;\Theta, \Lambda)= f(\mathbf{s}_{uij};\Lambda) \, \cdot \, \mathcal{L}_{\text{BPR}}(u,i,j;\Theta).
\label{eq:Importance-BPR}
\end{equation}

In our realization of the framework, $f(\mathbf{s}_{uij};\Lambda)$ is specified by the following equations:
\begin{equation}
    \begin{aligned}
        \mathbf{z}_{uij} &=\text{relu}(\mathbf{W}_1\cdot \mathbf{s}_{uij} +  \mathbf{b}_1) \,, \\
        w_{uij} &= \text{sigmoid}(\mathbf{W}_2\cdot \mathbf{z}_{uij} + \mathbf{b}_2)\,.
    \end{aligned}
\label{eq:MLP}
\end{equation}
Here $\mathbf{W}_1\in\mathbb{R}^{d\times 2d}$, $\mathbf{W}_2\in\mathbb{R}^d$, $\mathbf{b}_1\in\mathbb{R}^d$, and $\mathbf{b}_2\in\mathbb{R}$ are the learnable parameters $\Lambda$. $w_{uij}\in\mathbb{R}$ is the generated weight assigned to the triplet $(u,i,j)$. We note that applying a more carefully-tuned network as the weight generator may improve the performance, but this is not the focus of this work. Thus, we use a simple two-layer Multilayer Perceptron (MLP) to generate the weights.

A crucial question is how to decide whether a given triplet is important so that we can construct an informative input state $\mathbf{s}_{uij}(\Theta)$ for the weight importance generation function. We argue that generating weights by considering the training triplet in isolation is less effective than taking a broader perspective. We consider the triplet importance from another angle: we aim to measure how well the positive item and negative item match the principle interests of a certain user. If the positive item is close to the user's interest and the negative item is far from it, such a training triplet should have a high confidence in its truthfulness. As a result, we intend to assign this triplet a higher weight to emphasize its significance. Otherwise, we may downweigh the triplet.
\begin{figure*}[t]
    \centering
    \includegraphics[width=\linewidth]{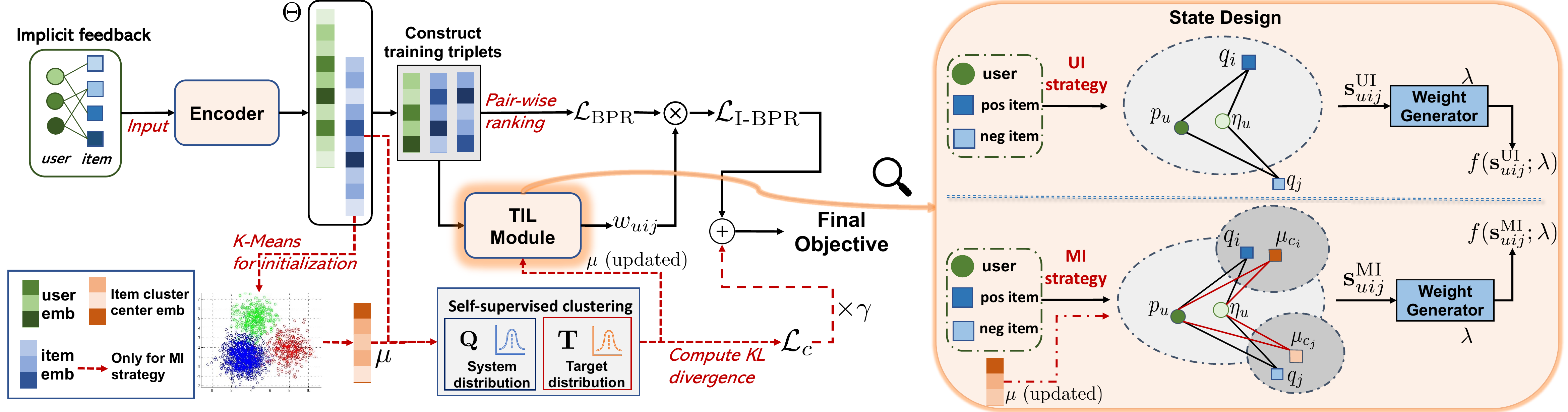}
    \vspace{-3mm}
    \caption{The architecture of the whole framework and the two strategies we proposed for capturing the influential factors for the importance score generation. On the right, each grey ellipse indicates an item community. The red lines in the ``State Design'' refer to the new relations we captured in the Multi-Interest Strategy through modeling the fine-grained user interests.}
\vspace{-5mm}
\label{pic:strategy}
\end{figure*}
To represent the user interests, we propose two strategies: \textbf{Uni-Interest} and \textbf{Multi-Interest}.



\subsubsection{Uni-Interest Strategy}
Intuitively, all the accessed items together can represent a user's primary interest. 
Thus an aggregation of these item embeddings may be a comprehensive description to illustrate the principle user interest:
\begin{equation}
    \bm{\eta}_u = \mathop{\text{aggregation}}\limits_{v \in \mathcal{S}_u^{+}} \bm{q}_v \,.
\label{eq:emb_avg}
\end{equation}
Here, we use an intuitive choice to conduct an arithmetic average as the aggregation function, where $\bm{\eta}_u = \frac{1}{|\mathcal{S}_u^{+}|} \sum_{v \in \mathcal{S}_u^{+}} \bm{q}_v$. 
Notice that we do not intend to explore other choices here, such as the attention mechanism, since this is not the main research point in this work.
Then we can measure the proximity of the positive item embedding $ \bm{q}_i $ to $ \bm{\eta}_u $, and the proximity of the negative item embedding $ \bm{q}_j $ to $ \bm{\eta}_u $. These values provide guidance as to the extent to which the items in the triplet align with the principle user interest. We design $ \mathbf{s}_{uij} $ to incorporate both the distances between the user and item embeddings and the distances between the item embedding and the principle user interest representation $ \bm{\eta}_u $:
\begin{equation}
\begin{aligned}
    \mathbf{s}_{uij}^{\text{UI}} =& ( \underbrace{\bm{q}_i \odot \bm{p}_u}_{\text{Term A}} + \underbrace{\bm{q}_i \odot \bm{\eta}_{u}}_{\text{Term B}}) ||\\
    & \, ( \underbrace{\bm{q}_j \odot \bm{p}_u}_{\text{Term C}} + \underbrace{\bm{q}_j \odot \bm{\eta}_{u}}_{\text{Term D}}) \,.
\end{aligned}
\label{eq:state-UI}
\end{equation}
Here ``$||$'' represents the concatenation, and ``$\odot$'' denotes element-wise multiplication. Terms A and C measure the user preferences for the positive item and the negative item based on the current embeddings. Terms B and D provide extra information about the extent to which the positive and negative items align with the principle user interest. Since Terms A and B reflect two distinct ways of representing the positive item's closeness to the user's preference, we use a summation to aggregate both information. We do the same operation on the negative item side for Terms C and D.

Any other operation that is capable of expressing the relationship between embeddings, such as the element-wise Euclidean distance, can be employed to replace the element-wise multiplication. 
We pick ``$\odot$'' since the multiplication operation is frequently employed in recommendation to compute the similarity of embeddings. 


\subsubsection{Multi-Interest Strategy}
\label{sc:MIS}
In the real world, a user may have several major interests. Thus, using a single embedding $\bm{\eta}_u$ to represent the full spectrum of a user's preference may be insufficient. To address this, we apply a clustering method to aggregate items that exhibit similar properties. 
Each cluster center now represents a more fine-grained user interest. In order to obtain these cluster center embeddings, we first obtain a cluster id $\bm{c}\in\mathbb{R}^{|\mathcal{V}|}$ for each item through some clustering algorithm, which will be elaborated in the following section. We first focus on how to construct the state vector $\mathbf{s}_{uij}$  given the cluster information. Assuming there are $K$ clusters, the embedding of the $k^{th}$ ($1\leq k\leq K$) cluster center $\bm{\mu}_k$ is the average embedding of those items belonging to the $k^{th}$ cluster, computed in a similar fashion to Eq.~\eqref{eq:emb_avg}. Our design of $ \mathbf{s}_{uij} $ under this setting can thus be formulated as:
\begin{equation}
\begin{aligned}
    \mathbf{s}_{uij}^{\text{MI}} =& ( \underbrace{\bm{q}_i \odot \bm{p}_u}_{\text{Term A}} + \underbrace{\bm{q}_i \odot \bm{\eta}_{u}}_{\text{Term B}})+\alpha\cdot( \underbrace{\bm{\mu}_{c_i} \odot \bm{p}_u}_{\text{Term a}} + \underbrace{\bm{\mu}_{c_i} \odot \bm{\eta}_u}_{\text{Term b}}) \ ||\\
    & \, ( \underbrace{\bm{q}_j \odot \bm{p}_u}_{\text{Term C}} + \underbrace{\bm{q}_j \odot \bm{\eta}_{u}}_{\text{Term D}})+\alpha\cdot( \underbrace{\bm{\mu}_{c_j} \odot \bm{p}_u}_{\text{Term c}} + \underbrace{\bm{\mu}_{c_j} \odot \bm{\eta}_u}_{\text{Term d}}) \,,
\end{aligned}
\label{eq:state_MI}
\end{equation}
where the ``||'' and ``$\odot$'' are defined the same as for Eq.~\eqref{eq:state-UI}, while $\bm{\mu}_{\bm{c}_i}$ and $\bm{\mu}_{\bm{c}_j}$ represent the centers of the clusters that item $i$ and $j$ belong to, which can also be interpreted as the fine-grained user interest associated by item $i$ and $j$, respectively.
$\alpha$ is the scaling factor for balancing the existing terms and the newly added information.
Built upon our first strategy, Terms A, B, C, and D directly come from the state design for the Uni-Interest.
In order to capture the proximity between the user interests at different levels, we further add four terms: Terms a and c measure the user preferences for the fine-grained user interests associated by the positive item and the negative item; Terms b and d provide extra information about the extent to which the fine-grained user interest of those items align with the principle user interest. 
Since the state design in Eq.~\eqref{eq:state-UI} has proved to be effective, we follow the same formation to concatenate these terms.
The first row in Eq.~\eqref{eq:state_MI} refers to the proximity of user interests relevant to the positive item $i$, while the second row in Eq.~\eqref{eq:state_MI} refers to the proximity of user interests relevant to the negative item $j$.
Thus, we achieve offering a more precise signal for indicating the importance of the triplet $(u,i,j)$ considering the user interests at different levels. 

It is also worthy to mention that when item $i$ and $j$ locate in the same cluster, then $\bm{\mu}_{c_i}=\bm{\mu}_{c_j}$, which results in Term a and b being identical to Term c and d, respectively.
Consequently, the extra information added on the positive item side and the negative item side is equivalent, which is as expected. 
Additionally, we do not add the negative sign before Terms C/c and D/d to emphasize that the correlation between the output weight $w_{uij}$ to the values of terms associated with the positive item $i$ is the inverse of that to the values of terms associated with the negative item $j$. 
This is due to the fact that such correlation can be learned through the non-linear layers of the MLP, as described in Eq.~\eqref{eq:MLP}.

\vspace{-3mm}
\subsection{End-to-end Clustering}

We now demonstrate how to obtain the cluster center embeddings across items in our Multi-Interest Strategy demonstrated in Section~\ref{sc:MIS}. One naive method is to iteratively apply the K-means algorithm on the item embeddings to obtain the cluster centers. However, this clustering approach is not optimized jointly with our downstream objective and cannot train in an end-to-end fashion.


Thus, motivated by~\cite{WangPHLJZ19AttributeCluster, Bo0SZL020StructureCluster}, we use a self-supervised method to learn the cluster centers in an end-to-end fashion. Specifically, we treat the embeddings of item cluster centers $\Phi\in\mathbb{R}^{K\times d}$ as learnable parameters ($K$ is the number of clusters). The K-means algorithm is used only once globally on all item embeddings for initialization.

In order to represent the similarity between an item embedding $\bm{q}_i$ and the cluster center embedding $\Phi_k$ (the $k^{th}$ row of $\Phi$), we use a Student’s t-distribution~\cite{Maaten_student} to model the probability of assigning item $i$ to cluster $k$, (i.e., a soft assignment):
\begin{equation}
    \bm{Q}_{ik}=\frac{(1+||\bm{q}_i-\Phi_k||^2/\tau)^{-\frac{\tau+1}{2}}}{\sum\limits_{j=1}^K(1+||\bm{q}_i-\Phi_{j}||^2/\tau)^{-\frac{\tau+1}{2}}}\,.
\label{eq:distribution}
\end{equation}
Here $\tau$ is the temperature that 
controls the sharpness of the distribution. Each row of $\bm{Q}$ represents the probability of assigning item $i$ across $K$ clusters.



Since our goal is to obtain a good clustering of the items with good cluster cohesion within each cluster and separation between clusters, we optimize the clustering training process by learning more from items closer to the cluster center with high confidence assignments~\cite{Bo0SZL020StructureCluster}. Specifically, we add a power factor on the distributions in $\bm{Q}$ to emphasize the role of those items who are closer to cluster centers, so that optimizing the embeddings can make the items cohere to cluster centers. Therefore, we define the ``target'' $\bm{T}$ built upon $\bm{Q}$ as below:
\begin{equation}
    \bm{T}_{ik}=\frac{\bm{Q}_{ik}^2/\sum_i\bm{Q}_{ik}}{\sum\limits_{j=1}^K(\bm{Q}_{ij}^2/\sum_i\bm{Q}_{ij})} \,.
\end{equation}

Then the clustering loss is defined as a KL-divergence between the two matrices $\bm{T}$ and $\bm{Q}$:
\begin{equation}
    \mathcal{L}_{c} = KL(\bm{T}||\bm{Q})=\sum\limits_{i=1}^{|\mathcal{V}|}\sum\limits_{j=1}^K \bm{T}_{ij}\log\frac{\bm{T}_{ij}}{\bm{Q}_{ij}} \,.
\end{equation}

Through minimizing this KL-divergence, the learning procedure forces the item embeddings to cohere to the cluster center embeddings, thus achieving learning optimal cluster centers in an end-to-end way. 


It is easy to incorporate this end-to-end clustering technique within the Multi-Interest Strategy. The cluster id for an item $i$ can be represented as $\bm{c}_i=\argmax_t \bm{Q}_{it}$. Then the $\bm{\mu}_{\bm{c}_i}$ and $\bm{\mu}_{\bm{c}_j}$ in Eq.~\eqref{eq:state_MI} can be represented as $\Phi_{\bm{c}_i}$ and $\Phi_{\bm{c}_j}$, respectively. Besides, we can also use the soft assignment to obtain $\bm{\mu}_{\bm{c}_i}$, we leave it as a future design choice.


\subsection{The Overall Training Framework} 
Having described the design of the state $\mathbf{s}_{uij}$ in detail, we now focus on how to learn the whole model efficiently. 

For the \textbf{Uni-Interest Strategy}, the optimization problem can be formulated as:
\begin{equation}
    \Theta^*, \Lambda^*=\argmin\limits_{\Theta, \, \Lambda}\sum\limits_{u}\sum\limits_{i\in S_u^+}\sum\limits_{j\in S_u^-}\mathcal{L}_{\text{I-BPR}}(u,i,j;\Theta, \Lambda) \,.
\label{eq:L_ada_bpr}
\end{equation}

For the \textbf{Multi-Interest Strategy}, we redefine the model parameters containing the learnable embeddings of users, items, and item cluster centers as $\Gamma=[\Theta||\Phi]$, and then the overall training objective can be formulated as:
\begin{equation}
    \Gamma^*, \Lambda^*=\argmin\limits_{\Gamma, \, \Lambda}\sum\limits_{u}\sum\limits_{i\in S_u^+}\sum\limits_{j\in S_u^-}\mathcal{L}_{\text{I-BPR}}(u,i,j;\Gamma, \Lambda) +\gamma\cdot\mathcal{L}_{c}(\Gamma)\,,
\label{eq:L_ada_bpr_c}
\end{equation}
where $\gamma$ is the scaling factor on the clustering loss. 


Unfortunately, directly minimizing either of these objective functions, Eq.~\eqref{eq:L_ada_bpr} or Eq.~\eqref{eq:L_ada_bpr_c}, does not achieve the desired purpose of generating suitable adaptive weights. Since the weight-related term explicitly appears in the loss function, constantly decreasing the value of the generated weight is a straightforward way to reduce the loss. As a result, all generated weights have very small values or are set to zero. Such an optimization only works with a weight summation constraint (e.g., $\sum_{\langle u,i,j\rangle}f(\mathbf{s}_{uij};\Lambda)=1$), otherwise it cannot ensure the learned embeddings are meaningful. However, directly optimizing Eq.~\eqref{eq:L_ada_bpr} or Eq.~\eqref{eq:L_ada_bpr_c} with this constraint may run into a practical challenge of batch learning with a sum constraint over the entire data.

\vspace{-1mm}
\subsubsection{Solving through Bilevel Optimization}
To bypass the aforementioned problem, we adopt a bilevel optimization approach to decouple the learning of the weights from the learning of the recommendations. This technique has been widely used in the field of parameter learning~\cite{DBLP:journals/corr/abs-2101-04849, lambdaopt, LiuDJ19_meta}, with a theoretical guarantee for its convergence~\cite{DBLP:conf/nips/ShuXY0ZXM19}. 



Specifically, we use an outer optimization to learn the parameters of the weight generator module and an inner optimization to learn the model parameters of the core recommendation system. Thus, we modify the original optimization problem and the overall optimization procedure for the Uni-Interest Strategy becomes to:
\begin{equation}
\resizebox{0.48\textwidth}{!}{
\begin{math}
\begin{aligned}
    \min\limits_{\Lambda}\mathcal{J}_{\text{outer}}(\Theta^*(\Lambda)):=\sum\limits_{u}\sum\limits_{i\in S_u^+}\sum\limits_{j\in S_u^-}\mathcal{L}_{\text{BPR}}(u,i,j;\Theta^*(\Lambda))\\
    \text{s.t. } \Theta^*(\Lambda)=\argmin\limits_{\Theta}\mathcal{J}_{\text{inner}}(\Theta, \Lambda):=\sum\limits_{u}\sum\limits_{i\in S_u^+}\sum\limits_{j\in S_u^-}\mathcal{L}_{\text{I-BPR}}(u,i,j;\Theta, \Lambda)
\end{aligned}
\end{math}}
\label{eq:bl_adap_bpr}
\end{equation}

With this approach, we perform alternating optimization to learn $\Theta$ and $\Lambda$ in order to minimize the adaptive weighted BPR loss. For the inner optimization, we fix $\Lambda$ and optimize $\Theta$, while for the outer optimization, we fix $\Theta$ and optimize $\Lambda$.

Similarly, we use the same approach to optimize for the Multi-Interest Strategy as below:
\begin{equation}
\resizebox{0.48\textwidth}{!}{
\begin{math}
\begin{aligned}
    \min\limits_{\Lambda}\mathcal{J}_{\text{outer}}(\Gamma^*(\Lambda)):=\sum\limits_{u}\sum\limits_{i\in S_u^+}\sum\limits_{j\in S_u^-}\mathcal{L}_{\text{BPR}}(u,i,j;\Gamma^*(\Lambda))\\
    \text{s.t. } \Gamma^*(\Lambda)=\argmin\limits_{\Gamma}\mathcal{J}_{\text{inner}}(\Gamma, \Lambda):=\sum\limits_{u}\sum\limits_{i\in S_u^+}\sum\limits_{j\in S_u^-}\mathcal{L}_{\text{I-BPR}}(u,i,j;\Gamma, \Lambda)+\gamma\cdot\mathcal{L}_{c}(\Gamma)
\end{aligned}
\end{math}}
\label{eq:bl_adap_bpr_c}
\end{equation}

\subsubsection{Framework Gradient Approximation}
Most existing models use gradient-based methods for optimization, where the idea is that given a parameter setting at time $t$, the model parameters can be updated to time $t{+}1$ by performing a step in the opposite direction of the gradient of the loss function. Accordingly, we can speed up the training by using an approximation that has been validated in~\cite{DBLP:conf/wsdm/Rendle12, DBLP:conf/iclr/LiuSY19}. Thus, we can avoid the expensive computation for calculating the exact framework gradient. 

We consider Eq.~\eqref{eq:bl_adap_bpr} as an example, where we can obtain the following equation:
\begin{equation}
    \nabla_{\Lambda}\mathcal{J}_{\text{outer}}(\Theta^*(\Lambda))\approx\nabla_{\Lambda}\mathcal{J}_{\text{outer}}(\Theta-\alpha\nabla_{\Theta}\mathcal{J}_{\text{inner}}(\Theta, \Lambda)) \,.
\end{equation}
In this expression, $\Theta$ denotes the current model parameter, and $\alpha$ is the learning rate for one step of the inner optimization. 

We can define a proxy function to link $\Lambda$ with the outer optimization:
\begin{equation}
    \widetilde{\Theta}(\Lambda):=\Theta-\alpha\nabla_{\Theta}\mathcal{J}_{\text{inner}}(\Theta, \Lambda)
\end{equation}
In practice, we use two optimizers, $\text{OPT}_{\Theta}$ and  $\text{OPT}_{\Lambda}$, to update
$\Theta$ and $\Lambda$ iteratively for simplicity. We do the same thing for optimizing Eq.~\eqref{eq:bl_adap_bpr_c} by using two separate optimizers $\text{OPT}_{\Gamma}$ and  $\text{OPT}_{\Lambda}$.


\section{Experiments}

    

\begin{table*}[t]
\caption{The performance comparison of all methods on four backbones in terms of \textit{R@20 (Recall@20)} and \textit{N@20 (NDCG@20)} in percentage (\%). The best and the second best performing methods in each row are boldfaced and underlined, respectively. The TIL-MI method has a statistical significance for $p\leq 0.01$ compared to the best baseline method (labelled with *) based on the paired t-test.}
\vspace{1mm}
\centering
\resizebox{0.95\textwidth}{!}{
\begin{tabular}{c|c|c|c|ccc|cccc|cc}
\toprule
\textbf{Datasets} & \textbf{Backbone} & \textbf{Metric} & \textbf{BPR} & \textbf{AOBPR} & \textbf{WBPR} & \textbf{PRIS} & \textbf{TCE} & \textbf{RCE} & \textbf{\makecell{TCE\\-BPR}} & \textbf{\makecell{RCE\\-BPR}} & \textbf{TIL-UI} & \textbf{TIL-MI}\\
\cline{1-13}
\multirow{8}{*}{\textit{Amazon\_CD}}&\multirow{2}{*}{\textbf{MF}} & R@20 &9.99 & 10.03 & 10.01 & 10.32$^*$ & 8.90 & 9.70 & 9.03 & 9.71 & \underline{10.88} &  \textbf{12.03}\\
& & N@20 &5.81 & 5.83 & 5.92 & 6.04$^*$ & 5.03 & 5.69 & 5.36 & 5.73 & \underline{6.30} & \textbf{7.11}\\
& \multirow{2}{*}{\textbf{NeuMF}} & R@20 & 11.03 & 11.08 & 11.05 & \underline{11.93}$^*$ & 10.64 & 10.05 & 11.77 & 11.42 & 11.85 & \textbf{13.33}\\
& & N@20 &6.40 & 6.43 & 6.62 & \underline{7.22}$^*$ & 6.24 & 5.87 & 6.86 & 6.77 & 7.01 & \textbf{7.83}\\
\cline{2-13}
& \multirow{2}{*}{\textbf{MGCCF}} & R@20 & 13.80 & 13.86 & 14.01 & 14.24$^*$ & 13.34 & 13.16 & 13.82 & 13.94 & \underline{14.30} & \textbf{14.67}\\
& & N@20 &8.02 & 8.08 & 8.07 & 8.24$^*$ & 7.72 & 7.44 & 8.20 & 8.11 & \underline{8.43} & \textbf{8.68}\\
& \multirow{2}{*}{\textbf{LightGCN}} & R@20 & 13.40 & 13.44 & 13.52 & 13.77$^*$ & 12.94 & 12.86 & 13.52 & 13.45 & \underline{13.92} & \textbf{14.21}\\
& & N@20 &7.90 & 7.95 & 7.96 & 8.15$^*$ & 7.59 & 7.21 & 8.12 & 7.98 & \underline{8.34} & \textbf{8.61} \\
\cline{1-13}
\multirow{8}{*}{\textit{Gowalla}} &\multirow{2}{*}{\textbf{MF}} & R@20 &12.19 & 12.22 & 13.02 & 14.88$^*$ & 13.05 & 12.83 & 13.94 & 13.82 & \underline{15.80} & \textbf{16.42} \\
& & N@20 &7.76 & 7.80 & 7.92 & 9.49$^*$ & 8.06 & 7.82 & 8.53 & 8.25 & \underline{9.77} & \textbf{10.09}\\
& \multirow{2}{*}{\textbf{NeuMF}} & R@20 & 13.55 & 13.59 & 14.35 & 15.83$^*$ & 13.62 & 13.15 & 14.80 & 14.76 & \underline{16.89} & \textbf{17.32} \\
& & N@20 &8.47 & 8.52 & 8.60 & 9.23$^*$ & 8.42 & 8.37 & 8.96 & 8.87 & \underline{10.90} & \textbf{11.17}\\
\cline{2-13}
& \multirow{2}{*}{\textbf{MGCCF}} & R@20 & 15.75 & 15.84 & 15.83 & 16.35$^*$ & 15.94 & 15.76 & 15.93 & 15.80 & \underline{17.33} & \textbf{18.25}\\
& & N@20 &9.70 & 9.81 & 9.82 & 9.94$^*$ & 9.86 & 9.73 & 10.00 & 9.79 & \underline{10.24} & \textbf{11.06}\\
& \multirow{2}{*}{\textbf{LightGCN}} & R@20 & 17.73 & 17.75 & 17.78 & \underline{18.03}$^*$ & 17.82 & 17.72 & 17.99 & 17.82 & 18.01 & \textbf{18.61}\\
& & N@20 &11.16 & 11.21 & 11.18 & \underline{11.30}$^*$ & 11.24 & 11.12 & 11.21 & 11.15 & 11.27 & \textbf{11.62}\\
\cline{1-13}
\multirow{8}{*}{\textit{Last.fm}} &\multirow{2}{*}{\textbf{MF}} & R@20 &17.35 & 17.38 & 18.99 & 20.28 & 19.99 & 19.00 & 20.30$^*$ & 18.50 & \underline{20.70} & \textbf{21.63}\\
& & N@20 &10.47 & 10.52 & 11.43 & 12.07 & 11.91 & 11.46 & 12.15$^*$ & 11.13 & \underline{12.48} & \textbf{13.22}\\
& \multirow{2}{*}{\textbf{NeuMF}} & R@20 & 19.17 & 19.18 & 20.80 & 21.15 & 19.21 & 18.96 & 21.50$^*$ & 21.46 & \underline{22.86} & \textbf{23.15} \\
& & N@20 &11.63 & 11.67 & 12.15 & 12.26 & 11.78 & 11.65 & 13.07$^*$ & 12.99 & \underline{14.10} &  \textbf{14.21}\\
\cline{2-13}
& \multirow{2}{*}{\textbf{MGCCF}} & R@20 & 21.26 & 21.32 & 21.28 & 21.97$^*$ & 20.44 & 20.12 & 21.80 & 21.22 & \underline{23.14} &  \textbf{23.44}\\
& & N@20 &13.06 & 13.33 & 13.30 & 13.50$^*$ & 11.73 & 11.45 & 13.47 & 13.19 & \underline{14.19} &  \textbf{14.35}\\
& \multirow{2}{*}{\textbf{LightGCN}} & R@20 & 23.31 & 23.33 & 23.35 & 23.92$^*$ & 22.36 & 22.05 & 23.82 & 23.30 & \underline{24.14} &  \textbf{24.85}\\
& & N@20 &14.25 & 14.28 & 14.26 & 14.68$^*$ & 12.59 & 12.08 & 14.51 & 14.20 & \underline{14.82} &  \textbf{15.35}\\
\bottomrule
\end{tabular}}
\label{tab:overall-performance}
\vspace{-3mm}
\end{table*}

\begin{table}[t]
\centering
\caption{The statistics of datasets.}
\vspace{1mm}
\label{tab:data_statistics}
\begin{tabular}{lrrr}
\toprule
& Amazon\_CD & Gowalla & Last.fm  \\
\hline
\#User & 11,346 & 29,858 & 23,385\\
\#Item & 32,705 & 40,981 & 34,186 \\
\#Interac & 466,501 & 1,027,370 & 982,798 \\
Density & 0.126\% & 0.084\% & 0.123\% \\
\bottomrule
\end{tabular}
\vspace{-4mm}
\end{table}

\subsection{Datasets}
\label{sec:datasets}

We evaluate our proposed method on three real-world datasets from different domains and of varying sparsity: \textit{Amazon\_CD}, \textit{Gowalla}, and \textit{Last.fm}. \textit{Amazon\_CD} is adopted from the Amazon review dataset~\cite{HeM16AmazonData}. The \textit{Gowalla} dataset~\cite{ChoML11Gowalla} was collected worldwide from the \textit{Gowalla} website (a location-based social networking website) over the period from February 2009 to October 2010. The \textit{Last.fm} dataset~\cite{Bertin-Mahieux2011LastfmData} was collected from the Last.fm website, and contains the music listening records of users. In order to be consistent with the implicit feedback setting, for the datasets with explicit ratings, we retain any ratings no less than four (out of five) as positive feedback and treat all other ratings as missing entries. We follow common practice used in prior works~\cite{sun2020_mgcf, DBLP:conf/sigir/Wang0WFC19, he2020_lgcn} to filter out potential high variability users and items which have fewer than ten ratings. Table~\ref{tab:data_statistics} shows the data statistics.
The datasets and code are available at: \url{https://github.com/haolun-wu/TIL}.


\subsection{Evaluation Protocols}
We employ cross-validation to evaluate the proposed model. We split the user-item interactions into the training set, validation set, and testing set according to the ratios 8:1:1. The performance of all models is evaluated in terms of \textit{Recall@k} and \textit{NDCG@k}. \textit{Recall@k} indicates the coverage of true (preferred) items that appear in the top-$k$ recommended items. \textit{NDCG@k} (normalized discounted cumulative gain) is a measure of ranking quality~\cite{JarvelinK02ndcg}. All experiments are run 5 times with the same seed to control the data partition. The average results are reported in Table~\ref{tab:overall-performance}. 


\subsection{Methods Studied}

Since designing a novel recommendation model from scratch is not the focus of this paper, we build our TIL strategy upon the following four recommendation backbones, where \textbf{MF}~\cite{DBLP:journals/computer/KorenBV09} and \textbf{NeuMF}~\cite{DBLP:conf/www/HeLZNHC17} are MF-based; \textbf{MGCCF}~\cite{SunZMCGTH19_MGCCF} and \textbf{LightGCN}~\cite{he2020_lgcn} are GNN-based.
The reason we choose them is because the MF-based models are the most widely used in recommendation and have been approved to be efficient in real-world cases.
The GNN-based models are state-of-the-art, which can achieve competitive performance.

Our main contribution is to design a novel paradigm for triplet importance learning in implicit feedback.
We first present the results for vanilla \textbf{BPR}~\cite{RendleFGS2009_bpr}, which samples the negative items with equal probability and treats each triplet equally when constructing the training objective. 
We then compare with the following training schemes which only use implicit feedback.
We do not compare with methods using auxiliary information since such information (e.g., dwell time) is not always available in real-world scenarios.

\textit{Negative Item Sampling Methods}:
\begin{itemize}[leftmargin=*]
\item \textbf{WBPR}~\cite{DBLP:journals/jmlr/GantnerDFS12} assumes that the popular items that a user has not interacted with are more likely to be true negative items.
\item \textbf{AOBPR}~\cite{DBLP:conf/wsdm/RendleF14} improves BPR with adaptive sampling that promotes the selection of popular negative items. 
\item \textbf{PRIS}~\cite{DBLP:conf/www/Lian0C20} assigns larger weights to more informative negative samples based on an importance sampling design.
\end{itemize}

\textit{Positive Pair Re-weighting Methods}:
\begin{itemize}[leftmargin=*]
\item \textbf{TCE} and \textbf{RCE}~\cite{DBLP:conf/wsdm/WangF0NC21} adaptively prune noisy positive interactions during training based on the loss value.

\item \textbf{TCE-BPR} and \textbf{RCE-BPR} are our implementations of the above methods, replacing the original point-wise loss with a pair-wise ranking loss objective for a fair comparison.

\end{itemize}

\textit{Our Proposed Methods}:
\begin{itemize}[leftmargin=*]
\item \textbf{TIL-UI} learns the triplet importance through the Uni-Interest Strategy.
\item \textbf{TIL-MI} learns the triplet importance through the Multi-Interest Strategy with the end-to-end clustering.
\end{itemize}

\begin{figure}[t!]
\centering
\includegraphics[width=\linewidth]{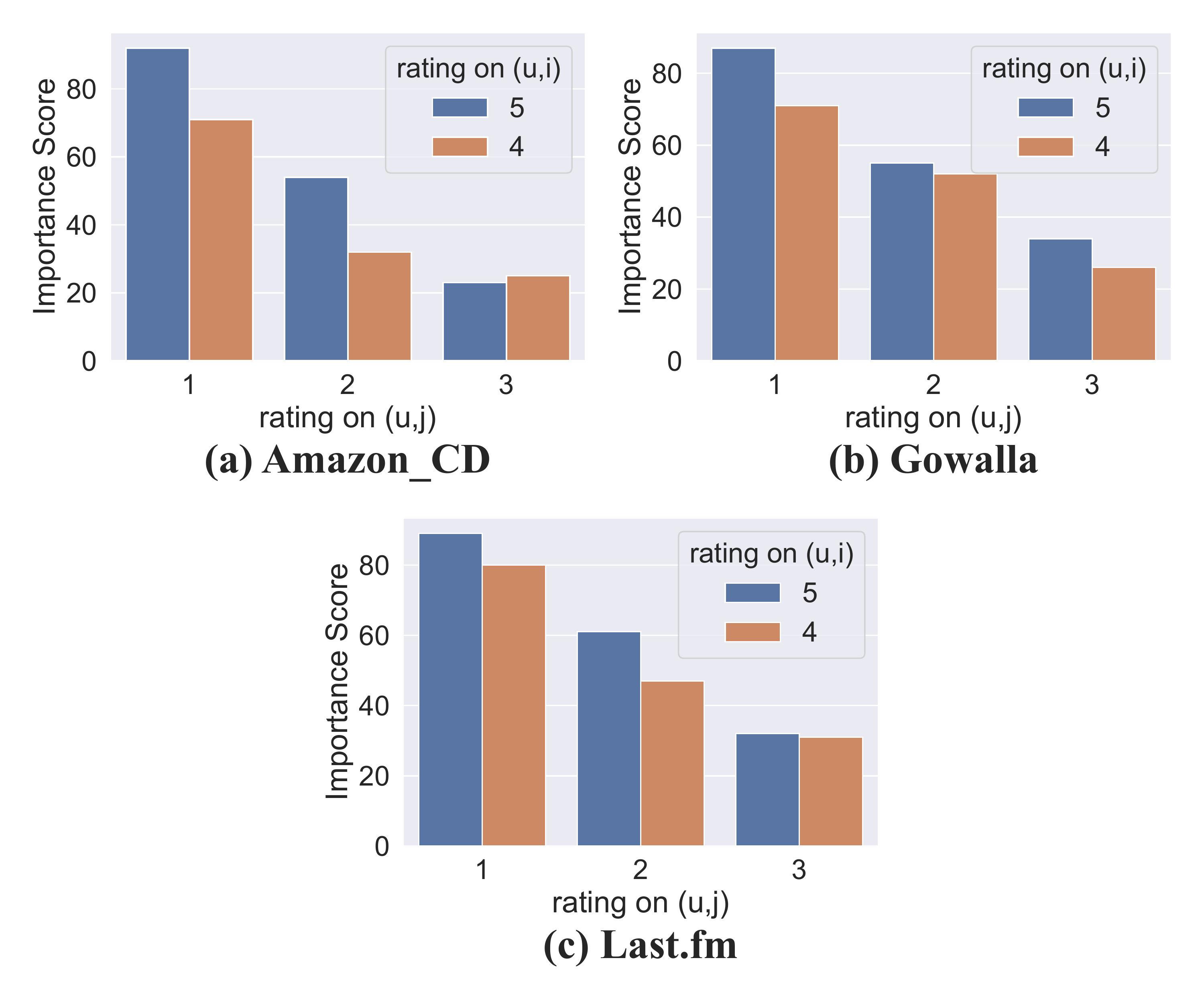}
 \vspace{-5mm}
\caption{The case study on the learned importance score.}
\label{pic:importance-score}
 \vspace{-6mm}
\end{figure}

\begin{figure*}[t!]
    \centering
    \includegraphics[width=0.99\linewidth]{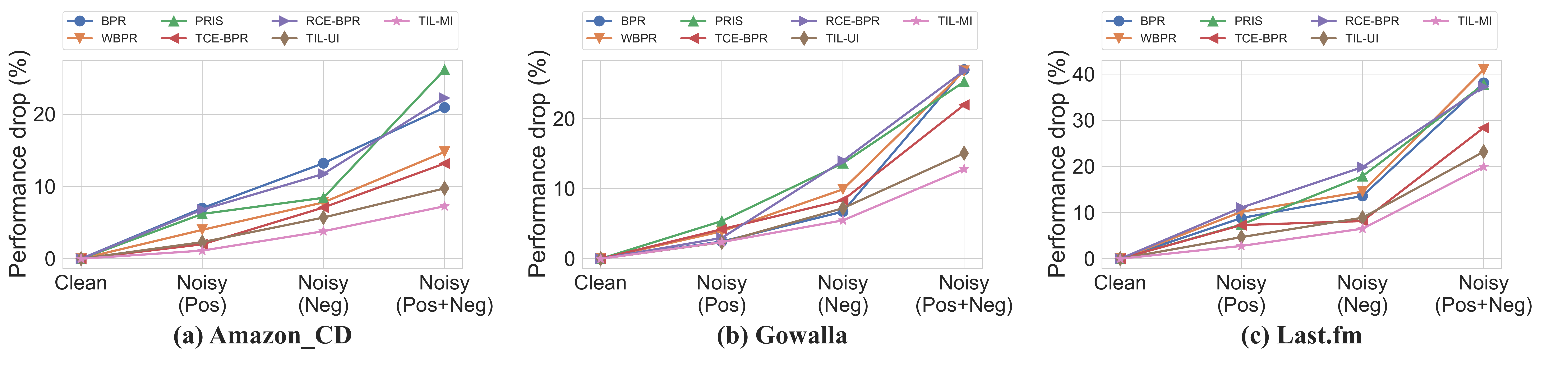}
     \vspace{-2mm}
    \caption{The robustness analysis on different methods with different noises added. As shown, our proposed methods, TIL-UI and TIL-MI, exhibit the least performance drop in all circumstances compared to all other methods studied.}
\label{pic:robustness}
\vspace{-4mm}
\end{figure*}

\begin{figure}[t!]
    \centering
    \includegraphics[width=\linewidth]{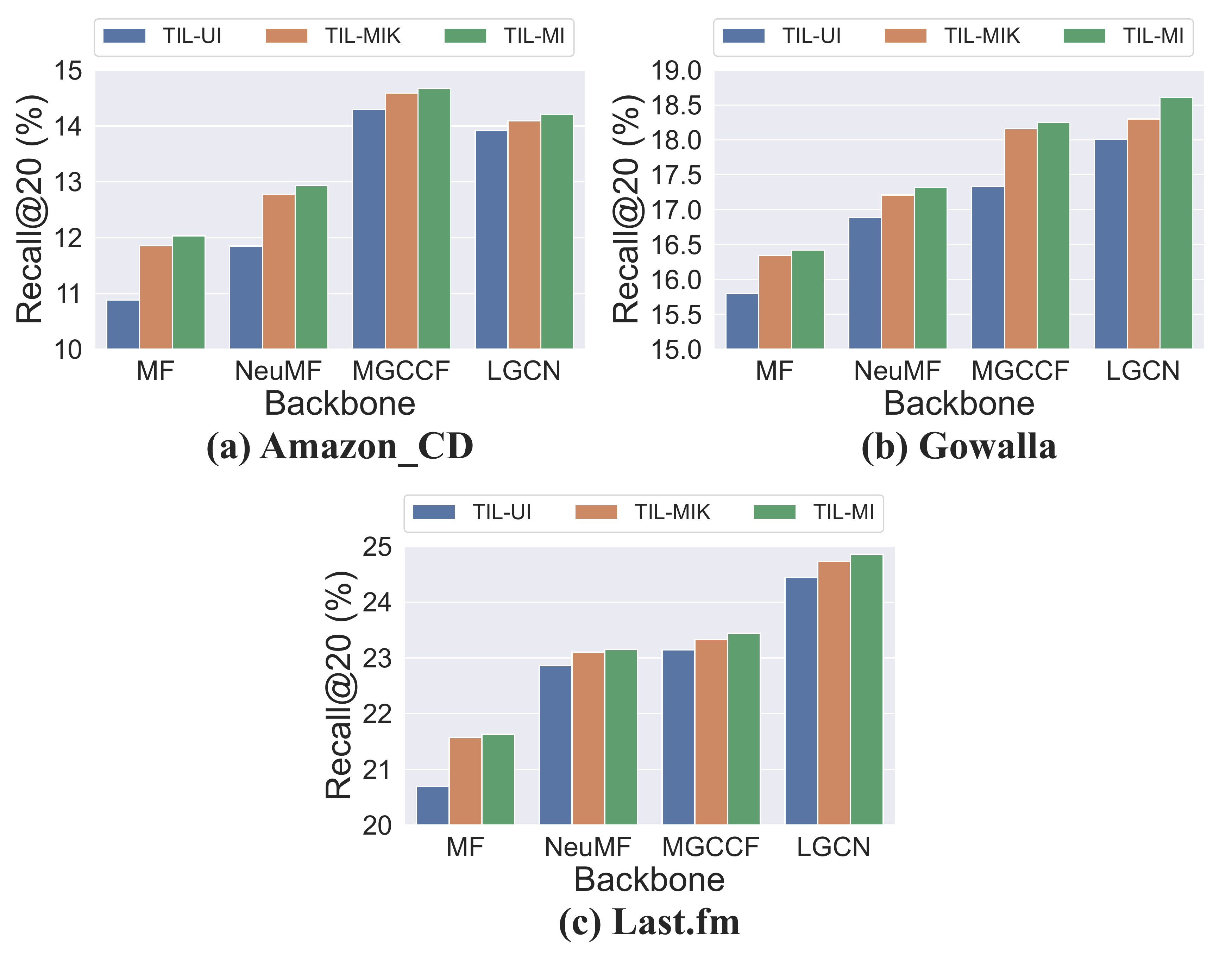}
    \vspace{-5mm}
    \caption{Model comparison for showing the effectiveness of the end-to-end clustering technique.}
\label{pic:model-comparison}
\vspace{-7mm}
\end{figure}

\begin{figure*}[t!]
    \centering
    \includegraphics[width=\linewidth]{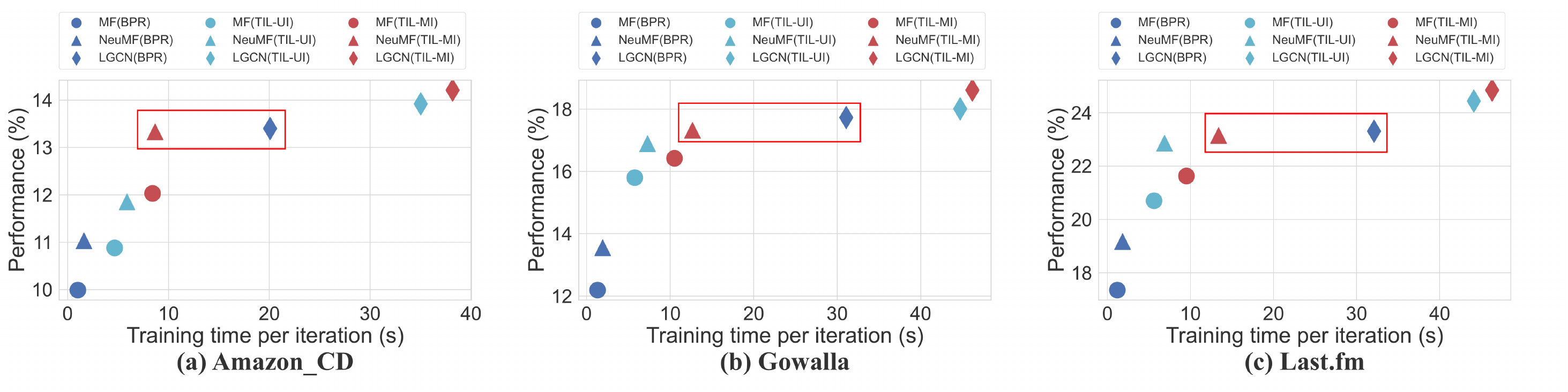}
     \vspace{-5mm}
    \caption{The analysis on the training efficiency versus the recommendation quality. We do not display the results of MGCCF in order to make the illustration more clear and easy to read. As shown in the red boxes, NeuMF with TIL-MI can achieve a very similar performance on all datasets compared to the vanilla LightGCN but with more than 50\% training time reduction.}
\vspace{-4mm}
\label{pic:time}
\end{figure*}

\subsection{Implementation Details}
We optimize all models using the Adam optimizer with the Xavier initialization procedure~\cite{DBLP:journals/jmlr/GlorotB10}. We fix the embedding size to 64 and the batch size to 5000 for all baseline methods and for our methods. When constructing the ranking loss objective, every positive item is associated with one sampled negative item for an efficient computation. Grid search is applied to choose the learning rate and the $L_2$ regularization coefficient $\lambda$ over the ranges $\{1e^{-4}, 1e^{-3}, 1e^{-2}\}$ and $\{1e^{-6}, 1e^{-5}, ..., 1e^{-2}\}$, respectively. 
The scaling factor $\alpha$ in Eq.~\eqref{eq:state_MI} is tuned from $\{1e^{-2}, 1e^{-1}, 1, 5, 10\}$. 
The scaling factor $\gamma$ in Eq.~\eqref{eq:bl_adap_bpr_c} is tuned from $\{1e^{-4}, 1e^{-3}, 1e^{-2}, 1e^{-1}\}$. 
The total number of training epochs is set to 3000 for all models with an early stopping design, i.e., premature stopping if Recall@20 on the validation data does not increase for 100 successive epochs. For the base models NeuMF, MGCCF, and LightGCN, we use the author-provided implementations. We use MGCCF and LightGCN with one graph convolution layer. In our proposed \textit{TIL-MI} method, the number of item clusters is set as 60 to balance the training efficiency and accuracy (we will discuss in later sections).
Since the embedding from the early training stage is less informative, we first pre-train the model without activating the clustering loss for obtaining meaningful embeddings.
(We set the number of epochs for the pre-training as 500.)
Then, the end-to-end clustering technique is applied to obtain the cluster membership of each item. To avoid instability during training, the cluster membership of each item is updated every 10 iterations during training. 
All experiments are conducted on GPU machines (Nvidia Tesla V100 32G).


\subsection{Overall Performance Comparison}

The overall performance comparison results are shown in Table~\ref{tab:overall-performance}.\\ 
\textbf{Observations about our method.} 
(\romannumeral1) Comparing with all the baselines that strive to make effective use of implicit feedback, our proposed TIL-UI and TIL-MI methods achieve the best performance on almost all the evaluation metrics. The superiority is consistent across the three datasets for four different backbone models. (\romannumeral2) TIL-MI consistently outperforms TIL-UI. This demonstrates that representing a user's preference as arising from a mixture of different interests is beneficial. (\romannumeral3) The performances of our proposed methods surpass all the sample re-weighting methods: TCE, RCE, TCE-BPR, and RCE-BPR. 
One reason is that all these four sample re-weighting methods assume the noise only comes from the positive samples, and thus treat all negative items equally, which is sub-optimal. Also, the positive sample weight only relies on the loss value, which might lead to reinforcing mistakes. (\romannumeral4) Compared to the negative sampling methods AOBPR, WBPR, and PRIS, the advantage of our method is apparent. This is likely because these negative sampling strategies can only select or re-weight more informative negative samples using handcrafted rules such as item popularity or the loss value given the sample, which might not be sufficient. Our designs adaptively learn the importance of each $(u,i,j)$ triplet via the personalized importance weight generation. 

\textbf{Other observations.} (\romannumeral1) TCE-BPR and RCE-BPR generally outperform TCE and RCE. The BPR pair-wise ranking loss used by the former appears to be more effective than the pointwise loss for the top-$k$ recommendation task with a ranking objective. (\romannumeral2) Although PRIS is a negative sampling method, in some cases, it shows competitive performance compared with alternative approaches such as AOBPR and WBPR. We believe that it benefits from the user-dependent design in its negative samplers that capture the user's preference for different item groups. By contrast, AOBPR and WBPR merely adjust the importance of a specific negative sample or adapt the probability of sampling it based on the item popularity in a user-independent way.

\subsection{Case Study and Analysis}
Due to the space limitation, we use the NeuMF as the backbone model to display the analysis results unless otherwise specified.
We consistently observe a similar trend in the following analysis across all the recommendation models we studied.

\subsubsection{Case Study on Triplet Importance} In order to investigate whether the importance scores learned through our TIL framework are meaningful, we conduct the following case study. From our training datasets with rating score information, we select the training triplets with rating score $\textit{rating}(u,i)\in\{4,5\}$ and rating score $\textit{rating}(u,j)\in\{1,2,3\}$ and report the final learned importance score by our model. The weights reported are normalized into 0 and 1 by the largest triplet weight and multiply by 100. 
As shown in Fig.~\ref{pic:importance-score}, if the item $i$ is highly likely to be a positive sample ($\textit{rating}(u,i)$ is high) and item $j$ is likely to be a negative sample ($\textit{rating}(u,j)$ is low), the triplet is regarded as important and a higher weight is assigned by our model. This empirical observation is consistent with our intended design.


\vspace{-0.1cm}
\subsubsection{Robustness Study on Noisy Dataset}
We believe that determining the significance of data can also help the model maintain its robustness when noisy samples are introduced. This is because if a model is capable of learning the unique importance of data during training, it should also be able to identify and downweigh those noisy samples. In this experiment, we examine the robustness of our methods by investigating the impact of introducing noise into the datasets. We design four kinds of datasets: \textbf{(\romannumeral1) Clean}: no modification of the datasets; \textbf{(\romannumeral2) Noisy-(Pos)}: we randomly introduce $c_u$ unobserved items as positive for each user $u$; \textbf{(\romannumeral3) Noisy-(Neg)}: we randomly make $c_u$ positive items that a user has interacted with as negative samples for that user; \textbf{(\romannumeral4) Noisy-(Pos+Neg)}: we combine the noisy data generation schemes in (\romannumeral2) and (\romannumeral3). We make $c_u$ user-dependent, setting it to $50\%$ of the total number of positive interactions for each user.
For each method training with the data setting (\romannumeral2), (\romannumeral3), and (\romannumeral4), we compute the performance (Recall@20) drop with respect to the same model training under the data setting (\romannumeral1). As shown in Fig.~\ref{pic:robustness}, our proposed methods have shown to be more robust against noisy data over other baselines. The advantage is more apparent when both the positive side and the negative side have introduced noise.

\subsubsection{Effectiveness of End-to-end Clustering} 

To demonstrate the efficiency of the end-to-end clustering technique in our TIL-MI model, we compare it to another model, named TIL-MIK, that iteratively updates the item cluster centers for the Multi-Interest Strategy using the K-means algorithm. The primary distinction between TIL-MIK and TIL-MI is that TIL-MIK does not learn cluster centers but instead uses the K-means results directly, whereas TIL-MI directly learns the embeddings and only uses K-means for initialization. For a fair comparison, we fix the number of clusters to 60 for both TIL-MIK and TIL-MI. 
We disable the item clustering module for the pre-training epochs to first obtain meaningful embeddings.
After pre-training, we update the cluster membership of each item in both models every 10 iterations. We choose Recall@20 as an evaluation metric and show the comparison of TIL-UI, TIL-MIK, and TIL-MI on three datasets with three different backbones. Based on the results shown in Fig.~\ref{pic:model-comparison}, we have two observations: (\romannumeral1) Along with TIL-MI, TIL-MIK also outperforms TIL-UI substantially in all scenarios. This again demonstrates the benefit of representing a user's preference as a composite of diverse interests. (\romannumeral2) TIL-MI consistently outperforms TIL-MIK. This confirms the advantage of the end-to-end clustering technique that directly learns the cluster center embeddings during the training procedure.

\subsubsection{Training Efficiency vs Recommendation Quality}
We investigate the trade-off between training efficiency and  model performance on the three public datasets. 
We present the analysis for BPR, TIL-UI, TIL-MI with three different backbone models: MF, NeuMF, and LightGCN. 
We do not show the MGCCF in the illustration in order to make the plots more clear and easy to read.
As shown in Fig.~\ref{pic:time}, we observe that by applying our proposed strategies, the performance (Recall@20) of simple base models can be greatly improved with reasonable computation overhead. 
For example, as shown in the red boxes in Fig.~\ref{pic:time}, a simple NeuMF with our designed TIL-MI training scheme \textit{($\redlozenge$ in red)} can achieve a very similar performance on all datasets compared to the vanilla LightGCN \textit{($\bluelozenge$ in blue)} but with more than 50\% training time reduction.
This strongly indicates the potential of applying our proposed methods on web-scale industrial datasets.



\vspace{-1mm}
\section{Conclusion}
In this work, we propose a novel training paradigm, TIL, to improve the training procedure of personalized ranking by adjusting the importance assigned to different triplets. We design a weight generation function to learn triplet importance which avoids enumerating different weights for individual triplets. We develop two strategies, Uni-Interest and Multi-Interest, to provide suitable inputs for the weight generation function based on how we model the user's interest. A bilevel optimization formulation is adopted to ensure the learned weights are meaningful, instead of all tending to zero values. 
Extensive experiments on three real-world datasets and four commonly used base recommendation models show that TIL significantly outperforms the SOTA methods for the top-$k$ recommendation task and is compatible with numerous backbone models.


\newpage
\bibliographystyle{ACM-Reference-Format.bst}
\balance
\bibliography{references}

\end{document}